\documentclass{aa_mod}

\usepackage{graphicx}
\usepackage{txfonts}
\usepackage{amsmath}
\usepackage{caption,subcaption}
\usepackage{placeins}
\usepackage{mathrsfs}
\usepackage{natbib}
\usepackage{xspace}
\usepackage{xcolor}
\usepackage{scalerel}
\usepackage[normalem]{ulem}
\usepackage{enumitem} 
\usepackage{multirow}
\usepackage{threeparttable, tablefootnote}
\usepackage{longtable}
\usepackage[colorlinks=true,linkcolor=blue,citecolor=blue, urlcolor=blue]{hyperref}
\bibpunct{(}{)}{;}{a}{}{,}
\usepackage{multirow}

\newcommand{\Jzeroun}{QJ~0158$-$4325\xspace}

\newcommand{\rzero}{\ensuremath{R_{\rm 0}}}
\newcommand{\rref}{\ensuremath{R_{\rm MK11}}\xspace}
\newcommand{\rblr}{\ensuremath{R_{\rm BLR}}\xspace}

\newcommand{\rblrref}{\ensuremath{R_{\rm BLR_{\rm MK11}}}\xspace}

\newcommand{\ve}{\ensuremath{v_{\rm e}}\xspace}
\newcommand{\thetae}{\ensuremath{R_{\rm E}}\xspace}
\newcommand{\kms}{km\ensuremath{\cdot} s\ensuremath{^{-1}}\xspace}
\newcommand{\kmsmpc}{km\, s\ensuremath{^{-1}}\, Mpc\ensuremath{^{-1}}\xspace}
\newcommand{\invday}{days\ensuremath{^{-1}}\xspace}
\newcommand{\fml}{\ensuremath{f_{\rm M/L}}\xspace}
\newcommand{\fblr}{\ensuremath{f_{\rm BLR}}\xspace}
\newcommand{\sigmadrw}{\ensuremath{\sigma_{\rm DRW}}\xspace}
\newcommand{\meanm}{\ensuremath{\left<M\right>}\xspace}
\newcommand{\solarm}{\ensuremath{M_{\odot}}\xspace }
\newcommand{\tiretsigma}{\ensuremath{-\sigma}\xspace }
\newcommand{\deltat}{\ensuremath{\Delta t_{\rm AB} }\xspace }

\definecolor{green}{HTML}{4CFF33}

\begin{document} 

\title{Constraining quasar structure using high-frequency microlensing variations and continuum reverberation}


   \author{E. Paic\inst{\ref{epfl}}, G.Vernardos \inst{\ref{epfl}}, D. Sluse\inst{\ref{liege}}, M. Millon\inst{\ref{epfl}},  F. Courbin\inst{\ref{epfl}}, J.H. Chan \inst{\ref{epfl}}, V. Bonvin\inst{\ref{epfl}}} 

   \institute{Institute of Physics, Laboratory of Astrophysics, Ecole Polytechnique 
F\'ed\'erale de Lausanne (EPFL), Observatoire de Sauverny, 1290 Versoix, 
Switzerland \label{epfl}\goodbreak \and
STAR Institute, Quartier Agora - All\'ee du six Ao\^ut, 19c B-4000 Li\`ege, Belgium \label{liege} \goodbreak}

\date{Received July 16th; accepted October 8th, 2021}
 
  \abstract{Gravitational microlensing is a powerful tool to probe the inner structure of strongly lensed quasars and to constrain parameters of the stellar mass function of lens galaxies. This is done by analysing microlensing light curves between the multiple images of strongly lensed quasars, under the influence of three main variable components: 1- the continuum flux of the source, 2- microlensing by stars in the lens galaxy and 3- reverberation of the continuum by the Broad Line Region (BLR). The latter, ignored by state-of-the-art microlensing techniques, can introduce high-frequency variations which we show carry information on the BLR size. 
  We present a new method which includes all these components simultaneously and fits the power spectrum of the data in the Fourier space, rather than the observed light curve itself.
  In this new framework, we analyse COSMOGRAIL light curves of the two-image system \Jzeroun known to display high-frequency variations. Using exclusively the low frequency part of the power spectrum our constraint on the accretion disk radius agrees with the thin disk model estimate and previous work that fit the microlensing light curves in real space. However, if we also take into account the high-frequency variations, the data favour significantly smaller disk sizes than previous microlensing measurements. In this case, our results are in agreement with the thin disk model prediction only if we assume very low mean masses for the microlens population, i.e. \meanm = 0.01 \solarm. At the same time, including the differentially microlensed continuum reverberation by the BLR successfully explains the high frequencies without requiring such low mass microlenses. This allows us to measure, for the first time, the size of the BLR using single-band photometric monitoring, \rblr = $1.6^{+1.5}_{-0.8}\times 10^{17}$cm, in good agreement with estimates using the BLR size-luminosity relation.}


\authorrunning{Paic et al.}
 \maketitle
%

\section{Introduction}
\label{introduction}

There is a plethora of astrophysical and cosmological applications of strongly lensed quasars. The photometric variability of the multiple lensed images allows us to measure the time delays between arrival times of photons in the observer's frame and to measure cosmological parameters such as $H_0$ \citep[e.g.][]{refsdal64, wong20}. The lensing magnification offers an augmented view of quasar host galaxies by stretching the image of the regions in the immediate vicinity of the central supermassive black hole (SMBH) and therefore allows us to extend the study of co-evolution of galaxies and Active Galactic Nuclei (AGN) to otherwise inaccessible redshifts \citep{gebhardt00,peng06,ding17a, ding17b, ding21}. In the microlensing regime, photometric variations induced by stellar mass objects passing in front of the quasar images allow us to both study the fraction of mass under compact form in lensing galaxies and to dissect the structure of the central AGN on scales as small as parsecs or even light days, even at high redshifts \citep[see][for a general overview]{Schmidt2010}.

\begin{figure}
    \centering
    \includegraphics[width= \linewidth]{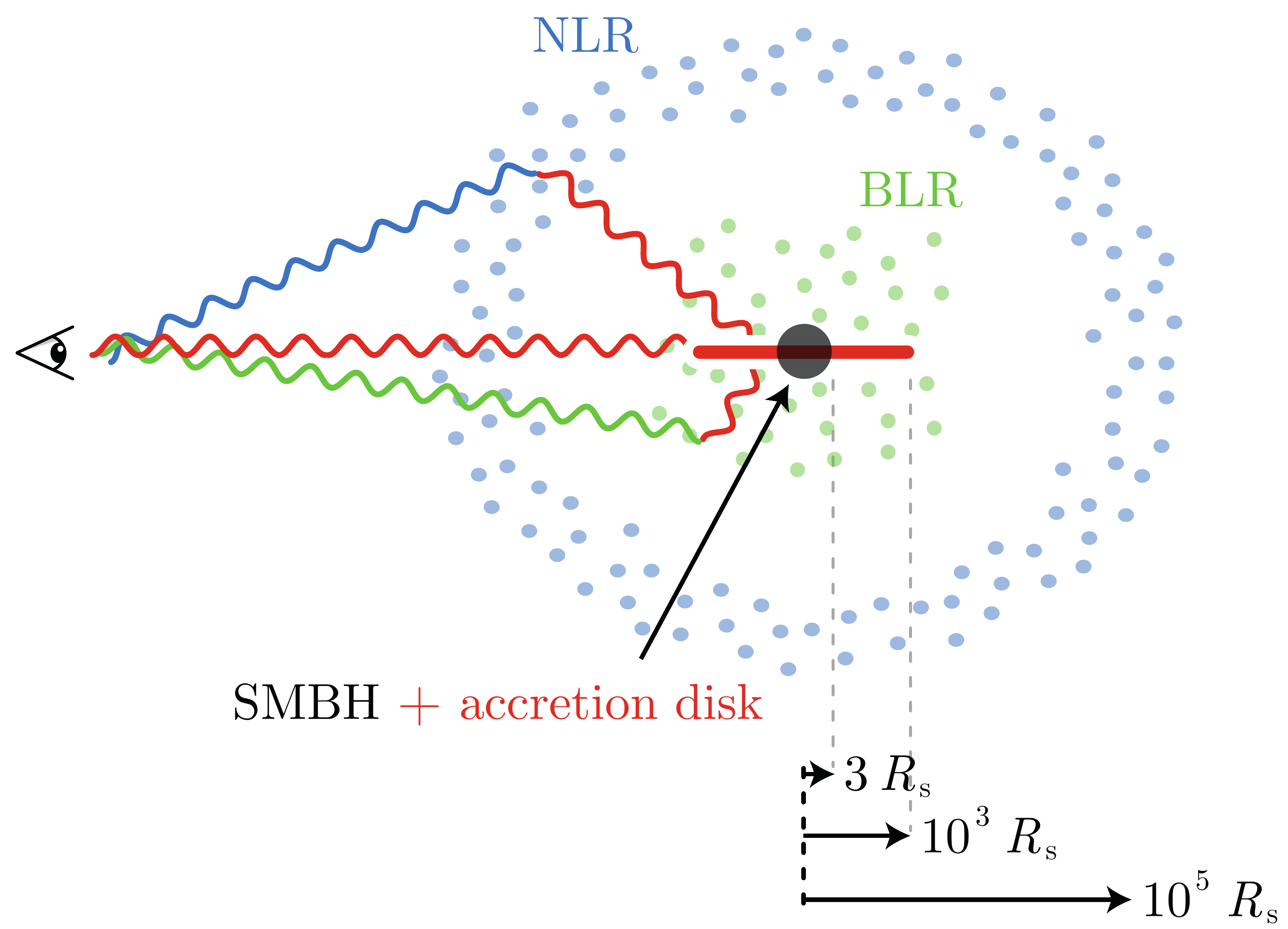}
    \caption{Schematic view of the relative location and size of the different regions of a quasar and illustration of the reverberation effect. $R_S$ is the Schwarzschild radius of the central Super Massive Black Hole. The continuum light from the central accretion disc (red) is reverberated both in the Broad Line Region (green) and in the Narrow Line Region (blue) that are much larger than the accretion disk and therefore much less affected by microlensing (see Sect.~\ref{reverberation_method}).
    }
    \label{schema_reverb}
\end{figure}

The bulk of quasar luminosity originates from the innermost regions containing the SMBH, surrounded by an accretion disk. Further out, clouds of ionised gas revolve around this central power engine and form the Broad and Narrow Line Regions (hereafter BLR and NLR), as illustrated in Fig.~\ref{schema_reverb} \citep[e.g.][]{urry95,elvis00}. The main difference between these two regions lies in their sizes, that imply different rotation velocities and therefore different widths of the observed spectral lines. As most of the energy in a quasar is generated from accretion processes in the central disk, it is essential to measure its size and energy profile. The latter is commonly assumed to follow the thin-disk model of \citet{shakura73} but active researches still test this model \citep[e.g.][]{edelson15,lobban20,li21}. Beyond the central accretion disk, the nature and dimensions of the BLR as well as its interaction with the host galaxy are still not fully understood \citep[e.g.][]{peterson06,czerny11,tremblay16}. Measuring the BLR size is therefore also of interest since it is related to the mass of the central SMBH \citep[e.g.][]{williams20}, hence constraining models of the inner structure of AGNs and of galaxy formation and evolution in general. 

Since BLR and accretion disks of quasars are generally smaller than $10^{-1}$ pc \citep{mosquera11}, these regions are not spatially resolved by any existing instrument\footnote{The Event Horizon Telescope: \url{https://eventhorizontelescope.org}, can resolve the accretion disk while the BLR can be resolved using the VLT but these are limited to a handful of nearby AGNs}, several techniques have been developed to infer the structure of the BLR and the accretion disk indirectly. The first measurements of accretion disk size were given by the flux-size estimate. It relies on the relation between the luminosity and the radius of the accretion disk, \rzero\xspace, given by the thin disk model and following \rzero\xspace $\propto L^{1/3}$ \citep{collin02, morgan10}.
Using this approach, \citet{mosquera11} predicted the radii of accretion disks in 87 strongly lensed quasars. Alternatively, continuum reverberation mapping, i.e. the measurement of the time-lag between different parts of the continuum, has been used to estimate the size of the accretion disk under the assumption of the lamp post model \citep[e.g.][]{krolik91,Chan20b}. Examples of such measurements of quasar accretion disks in non-lensed quasars can be found in \citep{mudd18,homayouni19,Yu2020}. As for the BLR size measurement, the reverberation mapping technique \citep{blandford82} relies on the time lag between those light rays that are coming straight from the accretion disk and those scattered (reverberated) by the BLR, as shown in Fig.~\ref{schema_reverb}. This method has been used to measure the size of the BLR and infer the mass of the SMBH through spectrophotometric monitoring \citep[e.g.][]{bentz09,du16,williams20,Kaspi21}.

A complementary and independent approach to study quasar structure is to use microlensing by stars in the lensing galaxy of strongly lensed quasars \citep{refsdal79}. In any given strongly lensed quasar, stars passing in front of the lensed images split the wavefronts of the incoming light, creating additional micro-images of the source separated by a few micro-arcseconds. The image splitting is not observable with existing instrumentation, but the resulting microlensing magnification is. In practise, the relative motion between observer, lens, microlenses, and source, induces a flickering of the macro-lensed, observable images. This flickering acts over time scales of weeks to years \citep[e.g.][]{mosquera11} and is a nuisance when measuring time delays \citep[e.g.][]{tewes13b,millon20} or macro-magnification ratios between the quasar images \citep{blackburne06}. However, because the variable micro magnification depends on the dimensions of the source, it also presents an opportunity to measure the size and energy profile of accretion disks \citep{Eigenbrod2008a} and to study quasar structure in general. Microlensing techniques are mainly sensitive to the half-light radius of the source and not so much on the shape of its light profile \citep{mortonson, vernardos19}. 

The apparent radius of an accretion disk is wavelength dependent since its inner region is hotter than its outskirts and therefore emits more energy \citep{shakura73}. As a consequence, the micro-magnification, which depends on the size of the source, depends on the wavelength of observation as well, even though microlensing is by nature achromatic. This leads to chromatic flux ratios between the observed images of a strongly lensed quasar, which have been identified in spectrophotometric monitoring data \citep[e.g.][]{Eigenbrod2008b}. Such chromaticity also enables the measurement of quasar accretion disks through single epoch multi-wavelength observations \citep{bate08} and although this method may be biased by the strength of the wavelength-dependent microlensing, it is a reliable way to measure the size of accretion disks \citep{bate19}. While in many strongly lensed quasars this method yields results in agreement with the thin disk theory \citep[e.g.][]{bate08,floyd09,mediavilla11,rojas14}, in other systems it was found that the thin disk model underestimated the size of the accretion disk by up to an order or magnitude \citet[e.g.][]{blackburne11,motta17,Bate2018,rojas20} which, according to \citet{cornachione20b}, would favour accretion disk temperature profiles shallower than predicted by the thin disk model. 


A second approach to measuring quasar structure with microlensing is to use the pair-wise difference light curves between quasar images, corrected for the time delay and macro magnification - often called 'microlensing light curves', as they are assumed corrected for intrinsic quasar variations by construction. Currently, most methods in use interpret such pair-wise difference light curves following the fitting technique introduced by \citet{kochanek04}. It consists of a Monte-Carlo analysis to compare huge amounts of simulated microlensing light curves ($\sim 10^{11}$), generated by varying a number of physical parameters on quasar structure, microlensing, and velocities, until a fit to the data is obtained. The main limitation of this approach is that, as the microlensing light curves get longer, their complexity grows due to the inclusion of more microlensing events, and consequently the amount of simulations required to fit the data rises drastically. Besides, this method assumes that microlensing occurs over long timescales, of the order of years, and focuses on the long-term effects while overlooking short-term variability. As a consequence, features shorter than $\sim1$ year have much less weight on the final inference than the longer features, resulting in a frequency filtering of the microlensing signal that may lead to an overestimate of the disk size\footnote{ \citet{dai10} show that accounting for a magnification offset due to contamination by the BLR of up to 40\% translates into shrinkage of the measured accretion disk  by up to 50\%. However, they do no actually consider the impact of reverberation on frequency content of the light curves.}. Measuring the disk size with this technique has been achieved by a number of authors \citep{morgan08,morgan12,morgan18,cornachione19}. By comparing their microlensing measurements with luminosity-based ones, \citet{morgan10} and \citet{cornachione19} pointed out a systematic discrepancy, which they explain by a possible shallower temperature profile of the disk than predicted by the thin disk model. Finally, a novel approach to microlensing light curve analysis was introduced by \citet{vernardos19}, employing machine learning to measure the accretion disk size. Such an approach has the potential of capturing both long and short term (low and high-frequency) variability in the signal, but it has not been applied to data yet.

High-frequency variations are visible in high cadence monitoring campaigns of strongly lensed quasars \citep[e.g.][]{millon20b}, potentially carrying valuable information on quasar structure, but their analysis is not possible with the light curve fitting method because of previously mentioned flaws. In addition, high-frequency signals can be introduced either by microlensing or can be partly due to  reverberation processes occurring between the inner accretion region and the BLR (e.g. see Fig.~\ref{schema_reverb}). The characteristics of the high-frequency variations depend on the relative sizes of the accretion disk and the BLR and lead to so-called 'microlensing-aided reverberation', first suggested by \citet[][]{sluse14}. In this work we consider this effect for the first time in the analysis of real data.

In order to study high-frequency variations, we introduce a new method relying on the Fourier power spectrum of the microlensing light curve, which allows one to characterise overall properties of the observed signal rather than any specific realisation of the light curve. Fitting the power spectrum enables us to investigate every time scale of variation, both in the high and low frequencies in a computationally tractable way. The method is applied to the light curve of \Jzeroun, already studied by \citet{morgan12} using the light curve fitting method, who estimated a disk size significantly larger than the one obtained by \citet{mosquera11} using a flux-based source size. As we show in our study, the power spectrum method demonstrates that part of the high-frequency variations in the microlensing light curve can be explained by continuum light being reverberated in the BLR. For the first time, we estimate the size of the BLR via microlensing-aided reverberation. 

The paper is organised as follows: Sect.~\ref{data} explains how the microlensing light curve of \Jzeroun is obtained, together with its power spectrum. Sect.~\ref{method} describes our new power-spectrum analysis approach. Sect.~\ref{results} explains the validation process of the method as well as the constraints obtained on the quasar's structure with and without taking into account the reverberation process. We conclude with a discussion of our results in Sect.~\ref{discussion}. Throughout this work we assume $\Omega_{\rm m} = 0.3$, $\Omega_{\Lambda} = 0.7$ and $H_{\rm 0} = 72$ \kmsmpc.

\section{Data \label{data}}

\begin{figure*}
    \centering
    \includegraphics[width=\linewidth]{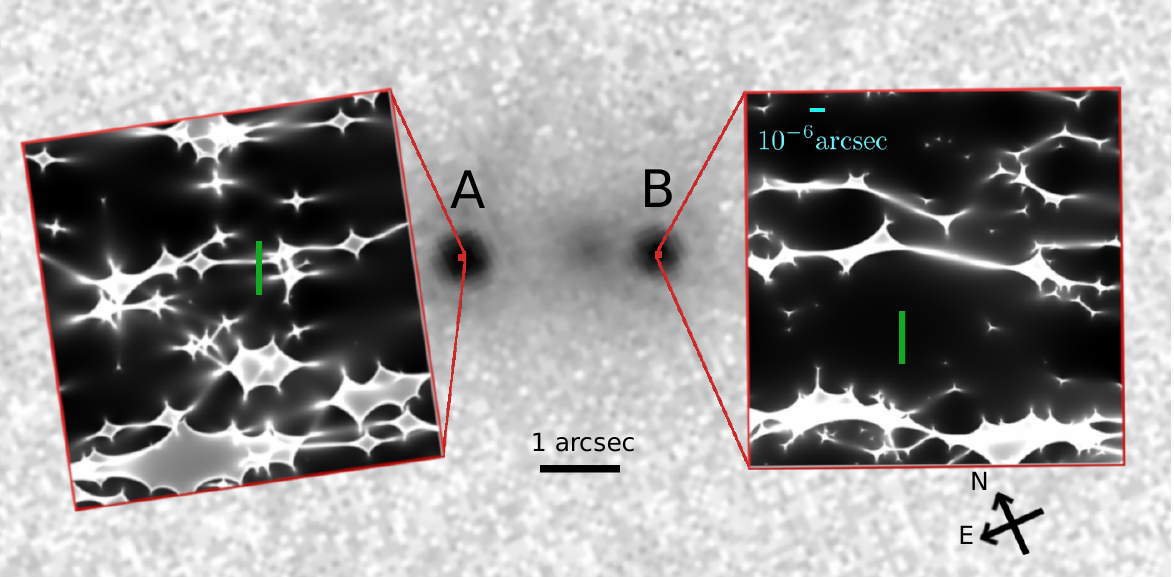}
    \caption{\Jzeroun observed with the Hubble Space Telescope in the F814W filter (program ID 9267; PI: Beckwith). The microlensing magnification maps (corresponding to \meanm= 0.3\solarm) are shown for each quasar image using the same colour scale, and are rotated with respect to the shear angle. The green line indicates a realisation of a trajectory of the source, the same in orientation and length for both maps drawn from the probability density function of \ve shown on Fig.~\ref{vdistrib}. }
    \label{lens}
\end{figure*}

\Jzeroun is a doubly-imaged quasar (see Fig. \ref{lens}) discovered by \citet{morgan99} that has been monitored for thirteen years by the Leonhard Euler 1.2m Swiss Telescope in the context of the COSmological MOnitoring of GRAvItational Lenses (COSMOGRAIL) program \citep{courbin05,eigenbrod05}. In \citet{millon20}, the light curves of this object were extracted following three main steps: first, the instrumental noise and the sky level were subtracted, then a Point Spread Function (PSF) estimated from nearby stars was fitted to each quasar image, and the flux was extracted at the image position with the MCS deconvolution algorithm \citep{MCS, cantale16}. This procedure allows one to extract the individual fluxes of the quasar images decontaminated from the light of the lensing galaxy \citep[see][for more details]{millon20}. The resulting light curves of \Jzeroun are shown in Fig.~\ref{lc_0158}.

\subsection{Microlensing light curve}

The observed light curve of a quasar image, $S_{\rm \alpha}(t)$, in magnitudes is the sum of the macro-magnification, $\mathcal{M}_{\rm \alpha}$, the intrinsic variations in each image, $V_{\rm \alpha}(t)$, and the microlensing magnification $m_{\rm \alpha}(t)$:
\begin{align}
\centering
    S_{\rm \alpha}(t)= \mathcal{M_{\rm \alpha}} + V_{\rm \alpha}(t) + m_{\rm \alpha}(t).
    \label{flux}
\end{align}
Without loss of generality, we can assume that the signal of image B of \Jzeroun is simply a time-shifted version of image A, i.e. $V_{\rm B}(t) = V_{\rm A}(t-\Delta t_{\rm AB})$, where $\Delta t_{\rm AB}$ is the time delay between the two images. Hence, the microlensing signal can be found by subtracting the observed light curves after correcting for the time delay and macro-magnification:
\begin{equation}
    S_{\rm A}(t)-S_{\rm B}(t-\Delta t_{\rm AB}) = m_{\rm A}(t) - m_{\rm B}(t) + \mathcal{M}_{\rm A} - \mathcal{M}_{\rm B}.
    \label{diff_ml}
\end{equation}
We note that we refer to this signal as 'microlensing' but it can well include a fraction of non-microlensed continuum light from the BLR, as will be shown later. This microlensing curve should therefore be seen as containing any 'extrinsic' variations, i.e. unrelated to the quasar instrinsic variability. We keep the macro magnification constant across the length of the light curve as we assume that it changes only at much longer time scales (see Table \ref{param_space} for the values used in this work)

As detailed in \citet{millon20}, the determination of \deltat is done by fitting the observed light curves with free-knot splines, implemented in the PyCS package \citep{pycs, millonjoss}. Such splines are piece-wise polynomials with the mean distance between two knots assigned by a parameter $\eta$ which controls the smoothness of the resulting fit. Free-knot splines allow the positions of the knots to be adjusted so that they capture both long and short features in the data being fitted. 
A single free-knot spline is fitted simultaneously to all light curves to model the intrinsic variation of the quasar while additional splines model the extrinsic (microlensing) variations in each light curve separately. A simultaneous fit of all instrinsic and extrinsic splines then allows us to adjust the time delays. 

\begin{figure*}
\centering
\includegraphics[width = 1.05\textwidth]{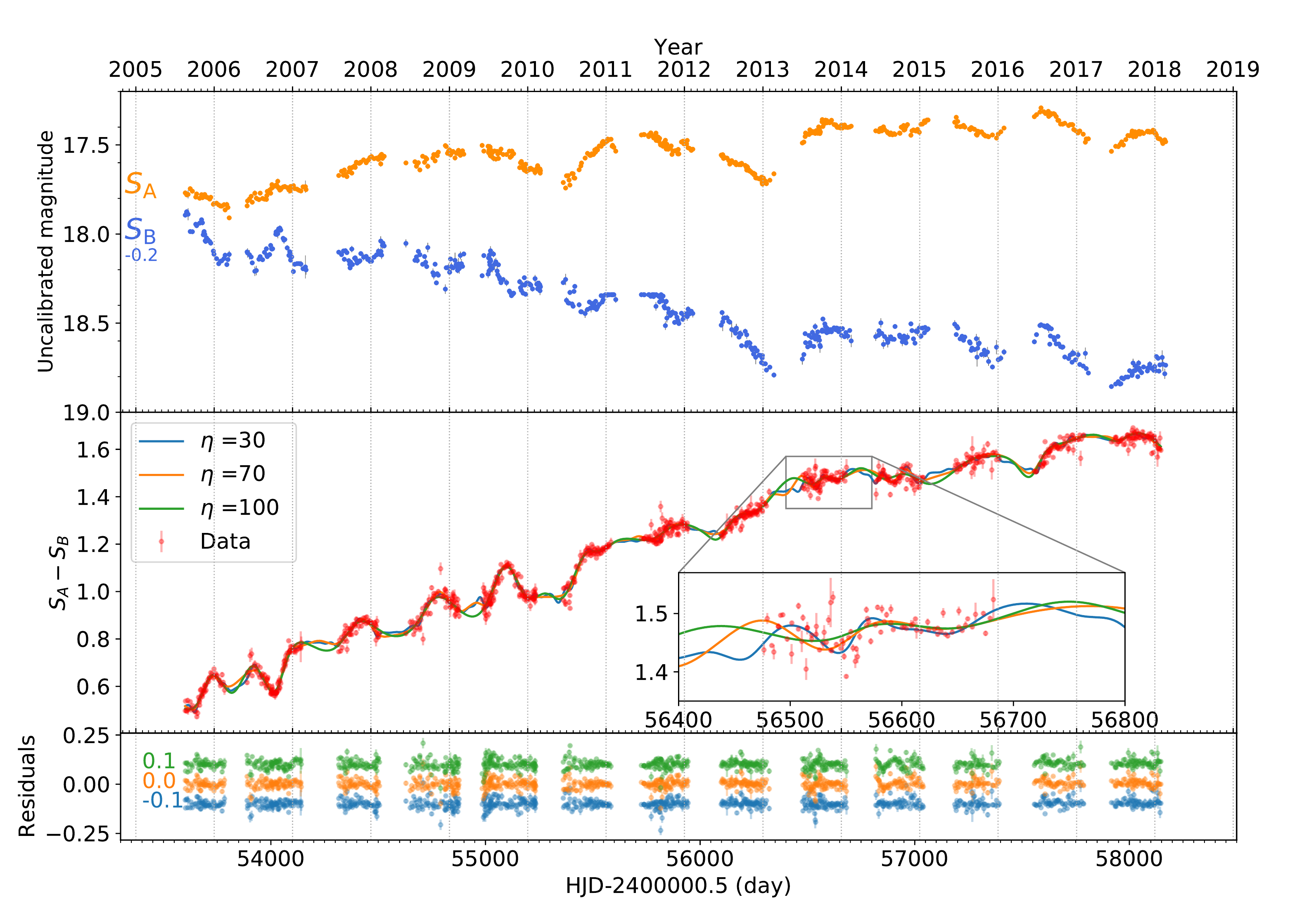}
\caption{\textit{Top}: COSMOGRAIL R-band light curves of images A and B of \Jzeroun over a period of 13 years. For clarity, the B curve has been artificially shifted upwards by 0.2 mag \citep[adapted from][]{millon20}. \textit{Middle}: Microlensing light curve (red) obtained from the observations using Eq.~(\ref{diff_ml}) with $\Delta t_{AB} = 22.7$ days \citep{millon20}, along with examples of spline fitting with different values of the $\eta$ parameter  (defined in Sect. \ref{psmlc}). \textit{Bottom}: Residuals of the three illustrative spline fits to the microlensing light curve. The number on the left indicates the artificial shifts applied for clarity purposes.
\label{lc_0158} }
\end{figure*}

For \Jzeroun, the time delay was found to be $\Delta t_{\rm AB} = 22.7 \pm 3.6$ days \citep{millon20} and the resulting microlensing light curve (Eq.~\ref{diff_ml}) is shown in the middle panel of Fig.~\ref{lc_0158}. We do not expect the uncertainty on the time delay to alter our constraints on the quasar structure because it is much smaller than the shortest time scale of interest - as discussed later, we focus on features in the differential light curve that are longer than 100 days for this study. The microlensing signal shows a steady rise throughout the period of observations, resulting in an overall increase of $\approx$1.2 mag, on top of which short modulations are observed within a single season; small scale variations are seen in the first 7 seasons (from 2005 to 2011) with a typical peak-to-peak amplitude of 0.1 magnitudes. Among the many lensed quasars monitored by the COSMOGRAIL project, very few exhibit such rich and diverse microlensing/extrinsic behaviour. \Jzeroun is therefore a promising test bench for investigating both high and low frequency variability.

\begin{figure}
\includegraphics[width = 1.1\linewidth]{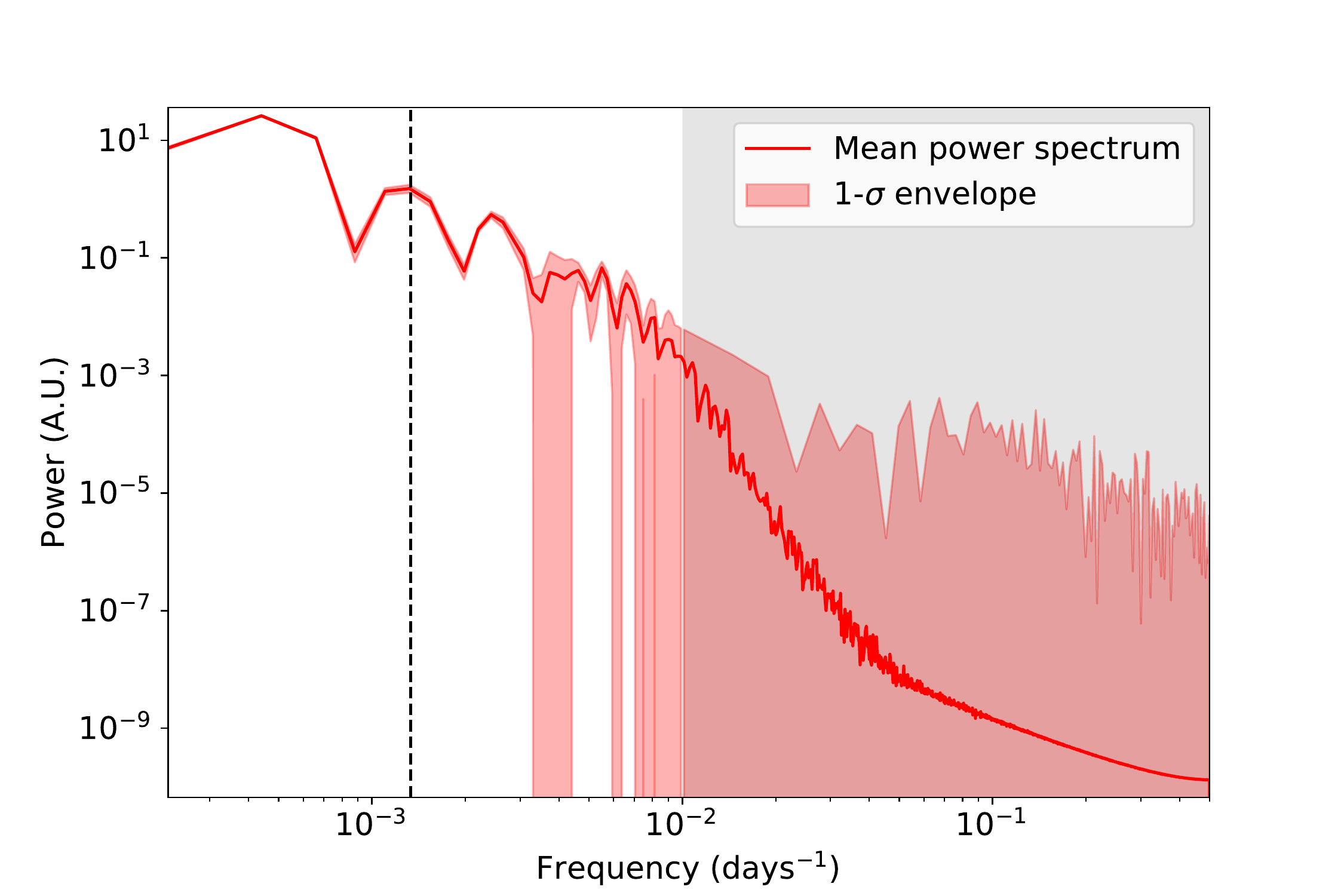}
\caption{Power spectrum of the observed microlensing light curve curve computed as the mean of the power spectra obtained for 1000 different realisations of photometric noise for every value of $\eta$ sampling the range [30:100] days with steps of 5 days used for the spline fitting parameter (see text). The 1\tiretsigma envelope is given by the standard deviation of the same set of power spectra.
The dashed black line marks the low/high-frequency boundary and the grey area indicates our adopted frequency limit of 10$^{-2}$\invday, below which photometric uncertainties dominate the data. 
\label{powerspectrum_data}}
\end{figure}

\subsection{Power spectrum of the microlensing light curve \label{psmlc}}

We represent the data in Fourier space in order to capture the high-frequency features that are missed by the light curve fitting method applied to \Jzeroun in \citet{morgan12}. The resulting power spectrum therefore holds the information across all the frequencies and allows us to treat high and low frequency signals simultaneously. In the following, we compute power spectra of both observed and simulated light curves using a standard Fourier transform. To tackle the so-called spectral leakage problem \citep{harris78}, i.e. the spurious broadening of spectral lines in frequencies for which the length of the signal is not a multiple of the corresponding period, we use a standard flat-top window function.

As shown in the upper panel of Fig.~\ref{lc_0158}, the light curves are not evenly sampled: within a season, two measurements may be separated by three or four days and throughout the entire curve, season gaps prevent the signal from being evenly sampled. The latter would in fact introduce a pattern in the Fourier transform that can be mistakenly interpreted as a periodic signal. To mitigate this, we interpolate the data through the season gaps using the continuous spline that resulted from the fitting technique to measure the time delays, as outlined above. This offers a flexible and model-independent way to fit time series. We set the sampling rate to one day and then compute the power spectrum.

The resulting power spectra have two main sources of uncertainty: on one hand, the photometric uncertainties of the raw data induce uncertainty in the very high frequencies corresponding to the sampling rate of the light curve (of the order of 1-10 days). On the other hand, the choice of the parameter $\eta$ can have a significant impact on the fitting of short time scale variations since, as illustrated in the middle panel of Fig.~\ref{lc_0158}, features shorter than $\eta$ are filtered out. The difference between underfitting and overfitting the data depends on the origin we attribute to a given short time scale variation and whether we want to discard it or not. Since we aim to use as few hypotheses as possible on the nature of these short variations, we don't make any assumption on the actual value of $\eta$ but rather consider a plausible range in order to estimate the uncertainty induced by this parameter. We define this range to [30:100] days in order be superior to the time sampling of the data while still capturing most of the high frequency features. 
The scatter of the residuals in the bottom panel of Fig.~\ref{lc_0158} shows that the selected range smoothly fits the data and most of the high-frequency variability is accounted for.

In order to quantify the uncertainty on the power spectrum induced by these effects, 1000 different realisations of photometric noise are produced for every value of $\eta$ sampling the range [30:100] days with steps of 5 days used for the spline fitting yielding a set of 14000 data like light curves. The power spectrum of the data that will be used further in this study is given by the mean and standard deviation of this set of light curves, shown in Fig.~\ref{powerspectrum_data}. As the extreme values of $\eta$ either overfit or underfit the light curve, the resulting uncertainty is conservative. We note that the relative uncertainties are negligible up to frequencies of $10^{-2}$ \invday. For higher frequencies, the power drops below 10$^{-3}$ and the relative uncertainties diverge because of the aforementioned two sources of uncertainties. Since the Einstein crossing time is around 18 years in the \Jzeroun system \citep{mosquera11}, we don't expect the light curve to contain features with time scales shorter than 100 days. As a result, there should not be much signal above the $10^{-2}$ \invday threshold and the power present in these frequencies is induced by photometric noise. Therefore, we exclude the frequencies above  $10^{-2}$ \invday (i.e. features shorter than 100 days) from the following analysis.

\section{Methods \label{method}}

In this section, the procedure of generating simulated light curves is described, as well as the way the simulated and observed power spectra are compared. The simulated variable flux of a quasar image, $F_{\alpha}(t)$, is assumed to be the combination of three components \citep{sluse14}: 
\begin{itemize}
    \item the intrinsic flux variability, $I(t)$, due to the stochastic emissions of the accretion disk,
    \item the microlensing magnification, $\mu(t)$, due to stars in the lens galaxy, and
    \item the flux arising from the BLR, $F_{\rm BLR}(t)$, which echoes the intrinsic variability of the continuum light of the accretion disk.
\end{itemize}
As the BLR is much larger than the accretion disk \citep[typically 10 times larger,][]{mosquera11}, microlensing of the resulting reverberated light is expected to be small\footnote{Although \citet{sluse12} showed that 10-20 per cent of the flux is typically microlensed, \citet{sluse14} found that this effect marginally impacts the variations in the microlensing light curve.}. Hence we have:
\begin{equation}
\centering
    F_{\rm \alpha}(t) = M_{\rm \alpha} \mu_{\rm \alpha}(t) I(t) + M_{\rm \alpha} F_{\rm BLR}(t),
    \label{fsluse}
\end{equation}
where $\mu_{\rm \alpha}$ is the time-dependent microlensing magnification and $M_{\rm \alpha}$ is the constant macro-magnification. Each component in this equation is separately described below.

\subsection{Intrinsic variability \label{intrinsic_variab}}

The variability of the accretion disk is commonly described by a damped random walk model \citep{kelly09, macleod10, ivezic13}. As the variability of \Jzeroun shows no major deviation from the standard quasar optical variability, we model it using such a model, parametrized by a characteristic timescale, $\tau_{\rm{DRW}}$, and amplitude, \sigmadrw, of the variations. We use the JAVELIN code presented in \citet{zu13} to create simulations of intrinsic light curves, designed to study the variability of quasars \citep{zu11}. A damped random walk consists of a Gaussian Process, $\mathcal{GP}$, with mean intensity $\overline{I}$ and covariance $Cov(\Delta t)$ between two moments in time separated by $\Delta t$ such that:
\begin{equation}
I(t) = \mathcal{GP}\left[\overline{I},Cov(\Delta t)\right],
\end{equation}
with the covariance given by :
\begin{equation}
Cov(\Delta t) = \sigmadrw ^2 \times \exp(-\left|\Delta t\right|/\tau_{\rm DRW}).
\end{equation}
Knowing $\tau_{\rm{DRW}}$ and \sigmadrw completely defines the $\mathcal{GP}$, from which different but equivalent realisations of the intrinsic light curve can be drawn.

\subsection{Microlensing variability \label{micro_variab}}

Magnification maps are used to simulate microlensing events produced by a given population of stars in the lens galaxy. In order to simulate a stellar population, we need to compute the values of the convergence, $\kappa$, the stellar surface density, $\kappa_{\rm *}$, and the shear, $\gamma$, at each image location from the smooth model of the lens galaxy mass distribution (i.e. macro-model, see Sect. \ref{results} and Table~\ref{param_space}). We then use a Salpeter initial mass function (IMF) to describe the stellar mass distribution around a given mean stellar mass \meanm. The maps are generated with the GPU-D software, which implements the direct inverse ray shooting method as described in \citet{vernardos15}, used in other microlensing studies as well \citep[e.g.][]{chan20}.


The characteristic scale of the magnification patterns created by such compact objects in the lens galaxy is their Einstein radius, \thetae, defined in the source plane as:
\begin{equation}
    \thetae = \sqrt{\frac{4G\left<M\right>}{c^2}\frac{D_{\rm ls}}{D_{\rm s} D_{\rm l}}},
    \label{thetaE}
\end{equation}
where $G$ is the gravitational constant, $c$ the speed of light, and $D_{\rm l}$, $D_{\rm s}$ and $D_{\rm ls}$ correspond to the angular diameter distances from the observer to the lens, from the observer to the source, and from the lens to the source, respectively.
The map dimensions are 8192 $\times$ 8192 pixels, corresponding to a physical size of 20\thetae $\times$ 20\thetae and a pixel size of $0.0024$\thetae. Fig.~\ref{lens} shows the configuration of the \Jzeroun lens system along with a realisation of magnification maps corresponding to the $\kappa$, $\kappa_*$ and $\gamma$ given by the smooth mass model at the given quasar image positions.

In order to study the magnification of a finite-sized source by a given caustic, we need to assume a light distribution, projected on the plane of the sky. The accretion disk of the source is assumed to be described by the thin-disk model \citep{shakura73}, in which a monochromatic light profile as a function of radius is given by:
\begin{eqnarray}
\centering
I_{\rm 0}(R) & \propto & [\exp(\xi) -1]^{-1} , \quad \mathrm{where}\\
\xi & = & \left(\frac{R}{\rzero}\right)^{3/4}\left(1-\sqrt{\frac{R_{\rm in}}{R}} \right)^{-1/4},  \nonumber 
\label{thin_disk}
\end{eqnarray}
with \rzero\xspace being the scale radius, i.e the radius at which the temperature matches the rest-frame wavelength of the observation assuming black body radiation, and $R_{\rm in} < R $ is the inner edge of the disk. Here we assume $R_{\rm in}$ = 0\footnote{$R_{\rm in}$ is very small compared to \rzero\xspace and shouldn't have an impact on the result since the half light radius $R_{1/2}$ remains mostly unchanged} and we explore a range of \rzero\xspace values that contains the value estimated by \citet{mosquera11}, i.e. \rzero\xspace$\approx 0.067 \times \thetae$ (in the case of \meanm=0.3\solarm), which is $\approx$15 pixels, on the maps that we use. 
To compute the magnification induced on the source, we need to convolve the magnification map with the light profile.



\begin{figure}
    \centering
    \includegraphics[width = \linewidth]{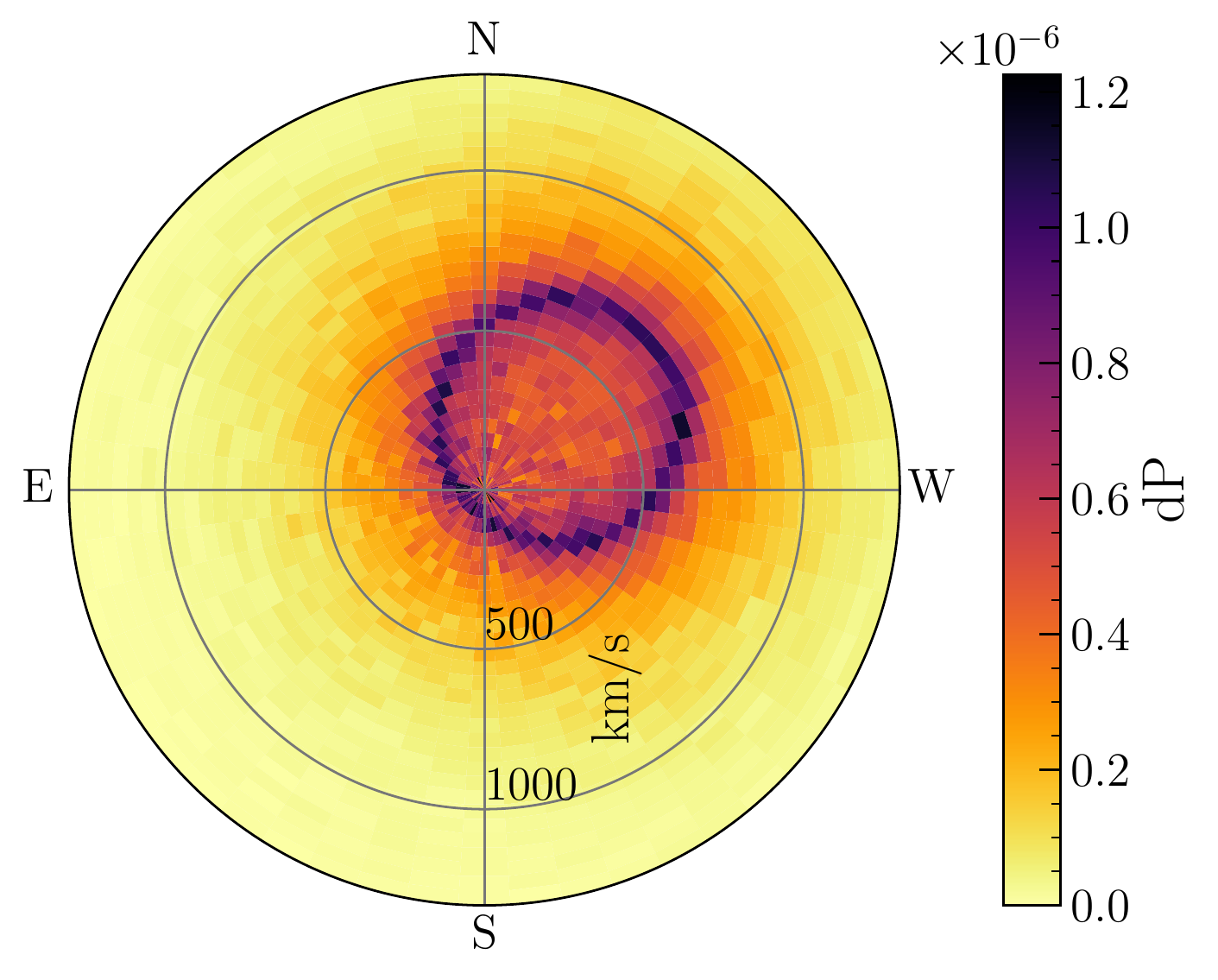}
    \caption{Probability density of the effective velocity, \vec{\ve}, in the source plane for \Jzeroun from Eq.~(\ref{ve}).}
    \label{vdistrib}
\end{figure}

The timescale of a microlensing event is set by the effective velocity in the source plane, \ve, which is the vectorial sum of the transverse velocities of the microlenses, $v_{\rm *}$, of the lens galaxy $v_l$, of the source, $v_s$, and of the observer, $v_o$. As described in \citet{neira20}, the direction of the microlens velocity is random, uniformly sampled in the [0;2$\pi$] interval. The magnitude of this velocity vector is given by: 
\begin{equation}
    v_{\rm *} = \sqrt{2}\, \epsilon\, \sigma_{\rm *},
\end{equation}
where $\epsilon$ is a factor depending on $\kappa$ and $\gamma$, assumed to be 1 \citep{kochanek04}, and $\sigma_{\rm *}$ is the velocity dispersion at the lens galaxy centre. The direction of $v_{\rm l}$ and $v_{\rm s}$ is random and their magnitude is drawn from a gaussian distribution with a given standard deviation, $\sigma_{\rm pec}(z)$ as a function of redshift. Therefore, these can be combined into a single normal variable $\vec{v_{\rm g}}$ with a random direction and a magnitude given by a gaussian with a standard deviation given by:
\begin{equation}
    \sigma_{\rm g}^2 = \left(\frac{\sigma_{\rm pec}(z_{\rm l})}{1+z_{\rm l}}\frac{D_{\rm s}}{D_{\rm l}}\right)^2+\left( \frac{\sigma_{\rm pec}(z_{\rm s})}{1+z_{\rm s}}\right)^2.
\end{equation}
The velocity of the observer is measured with respect to the cosmic microwave background velocity dipole: 
\begin{equation}
    \vec{v_{\rm o}} = \vec{v_{\rm CMB}} - (\vec{v_{\rm CMB}}\cdot \hat{z}) \hat{z},
\end{equation}
where \vec{v_{\rm CMB}} is the measured velocity vector with respect to the cosmic microwave background, and $\hat{z}$ the observer's line of sight. This component's magnitude and direction is computed using the position of the object on the plane of the sky.
Combining these terms, the effective velocity is: 
\begin{equation}
    \vec{v_{\rm e}} = \frac{\vec{v_{\rm o}}}{1+z_{\rm l}} \frac{D_{\rm ls}}{D_{\rm l}} -\frac{\vec{v_{\rm *}}}{1+z_{\rm l}}\frac{D_{\rm s}}{D_{\rm l}} + \vec{v_{\rm g}}.
    \label{ve}
\end{equation}
In the case of \Jzeroun, $z_{\rm l} = 0.317$ and $z_{\rm s}=1.29$ \citep{chen12}. All other relevant parameters for \Jzeroun are given in Table~\ref{param_space} and the resulting probability distribution of $\vec{v_e}$ from which the effective velocity in the source plane is drawn is shown in Fig.~\ref{vdistrib}. The probability density function in Fig.~\ref{vdistrib} is approximated by a gaussian kernel density estimator and then sampled through the inverse transform sampling method.  

\subsection{Reverberated variability \label{reverberation_method}}

As shown in \citet{sluse14}, delayed reverberation of the continuum light from the BLR can significantly alter the observed microlensing signal with modulations on short time scales. In the case of \Jzeroun, \citet{faure09} have shown that the \ion{Mg}{ii} as well as the \ion{Fe}{ii} spectral lines, both arising from the BLR, fall into the R-band used in this work. It therefore makes sense to consider continuum reverberation as a mechanism contributing to the observed light curves.

We can describe the reverberation component in Eq.~(\ref{fsluse}) as $F_{\rm BLR}(t)=\fblr r(t)$, where \fblr is the flux ratio between the line and the continuum and $r(t)$ is the reverberated flux. The latter can be computed as a convolution, $r(t) = \Psi(t,\tau)*I(t)$, between the intrinsic signal, $I(t)$, and $\Psi(t,\tau)$, a time-lagging transfer function that depends on the radius of the BLR through a corresponding time lag $\tau=R_{\mathrm{BLR}}/c$. Eq.~(\ref{fsluse}) then becomes:
\begin{align}
\centering
    F_{\rm \alpha}(t) &= M_{\rm \alpha} \mu_{\rm \alpha}(t) I(t) + M_{\rm \alpha} \fblr\left[\Psi(t,\tau)*I(t)\right]. 
\label{final_eq}
\end{align}
In this work we model the reverberation region as a diffuse ionised gas cloud with the geometry of a thin shell \citep{peterson93}, so that $\Psi(t,\tau)$ is a top hat kernel with an amplitude $A$ and a width equal to twice the assumed time lag $\tau$:
\begin{equation}
    \Psi(t, \tau) =
\left\{
	\begin{array}{ll}
		 A/\tau  & \mbox{if } 0  \leq t < 2\tau , \\
		0 & \mbox{ otherwise} .
	\end{array}
\right.
\label{transferfunction}
\end{equation}     
The values of the \fblr and \rblr parameters examined here are given in Table~\ref{param_space}. 

\subsection{Light curve simulation and fitting}
\label{sim}

\begin{figure} [!h]
\includegraphics[width = 1.0\linewidth]{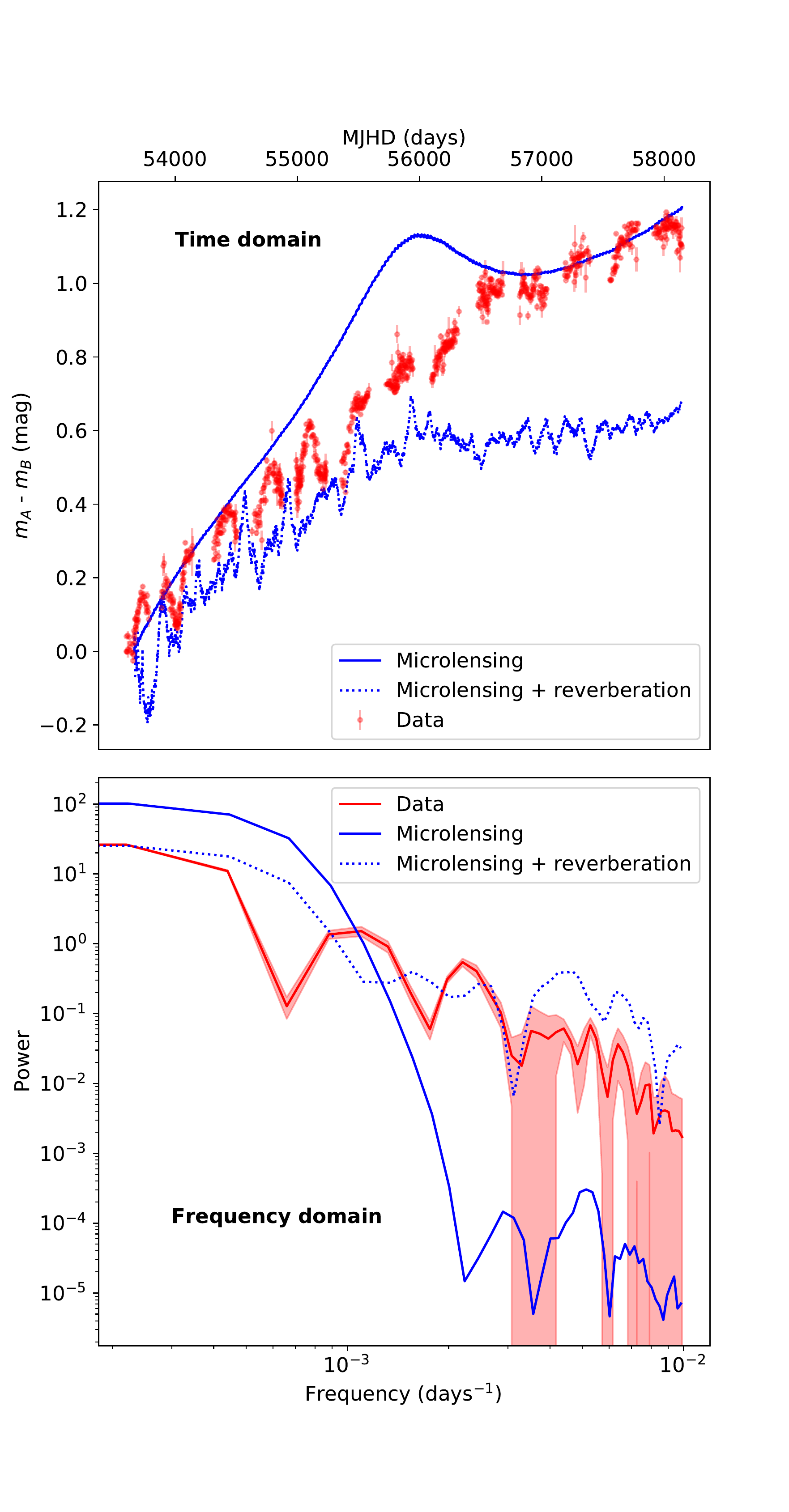}
\vspace{-1.5cm}
\caption{\textit{Top panel}: Example of a simulated light curve with and without reverberation (dotted and solid lines respectively). \textit{Bottom panel}: Corresponding power spectra. The curves and power spectra have been produced using \meanm = 0.3\solarm, \ve = 1236 \kms, \rzero = 0.5 \rref, \sigmadrw = 30 and \rblr=\rblrref. Adding reverberation clearly adds power to the high frequency part of the spectrum. \label{necessarynotsufficient}}
\end{figure}

Combining all the above model components, we are now able to simulate light curves for each quasar image using Eq.~(\ref{final_eq}). The free parameters are \meanm, \vec{\ve}, \rzero\xspace, \sigmadrw and \rblr, to which we will refer to as vector $\boldsymbol{\zeta}$. The final light curve to be compared to the data is obtained by dividing (subtracting) the flux (magnitudes) of pairs of simulated light curves for images A and B. Examples of simulated light curves with and without reverberation along with their corresponding power spectra are shown in Fig.~\ref{necessarynotsufficient}. 

For any given $\boldsymbol{\zeta}$, a batch of $10^5$ curves is created from the magnification maps and their power spectrum is computed. The mean $P_{\rm sim}(\omega)$ and standard deviation $\sigma_{\rm sim}(\omega)$ of the power spectrum in each frequency bin are then compared to the data using a chi-square statistic:
\begin{equation}
\centering
    \chi^2(\vec{\zeta}) = \frac{1}{N_{\rm \omega}} \sum_f \frac{\left(P_{\rm data}(\omega)-P_{\rm sim}(\omega)\right)^2}{\sigma_{\rm sim}(\omega)^2 + \sigma_{\rm data}(\omega)^2}, 
    \label{chi}
\end{equation}
where $P_{\rm data}(\omega)$ is the mean power spectrum of the data at the frequency $\omega$ and $\sigma_{\rm data}(\omega)$ its standard deviation shown in Fig. \ref{powerspectrum_data}. This can be turned into a likelihood through:
\begin{equation}
    \mathcal{L}(d|\boldsymbol{\zeta}) = \exp(-\chi^2(\boldsymbol{\zeta})/2).
\end{equation}

\begin{threeparttable}[htbp]
\caption{Fixed parameter values and free parameter ranges for the models of \Jzeroun used in this study (see Sect. \ref{method}). \label{param_space}}
\begin{tabular}{c c c c}
    & Name & Value & Unit \\
    \hline
    \hline
    \multirow{4}{*}{\begin{sideways}Intrinsic\end{sideways}}&&&\\
    & $\tau_{\rm DRW}$  \tnote{a}  & 810 & days \\
    & $\sigma_{\rm DRW}$ & [9 - 95] & Flux units \\
    &&&\\
    \hline
    \hline
    \multirow{22}{*}{\begin{sideways}Microlensing\end{sideways}}&&&\\ &\multicolumn{3}{c}{\textit{Lens mass model} \tnote{b}}\\
    &\fml & 0.9 & - \\
    &\meanm & [0.3, 0.1, 0.01] & \solarm\\
    & Corresponding \thetae & [3.41, 1.97,0.623] & $10^{16}$ cm\\
    & & & \\
    & \multicolumn{3}{c}{Image A}\\
    &$\kappa$ & 0.23 & -\\
    &$\gamma$ & 0.39 & - \\
    &$\kappa_{\rm *} /\kappa$ & 0.81 & - \\
    & $\mathcal{M}_{\rm A}$ & 2.24 & - \\ 
    &\multicolumn{3}{c}{Image B}\\
    &$\kappa$ & 0.72 & - \\
    &$\gamma$ &  1.03 & - \\
    &$\kappa_{\rm *} /\kappa$ & 0.92& - \\
    & $\mathcal{M}_{\rm B}$ & 0.84 & - \\
    \cline{2-4}
    &&&\\
    &\multicolumn{3}{c}{\textit{Effective velocity}}\\
    &$\sigma_{\rm pec}(z_{\rm l})$ \tnote{c} & 277 & \kms\\
    &$\sigma_{\rm pec}(z_{\rm s})$  \tnote{c}& 248 & \kms\\
    &$v_{\rm CMB}$ \tnote{d}&328 &\kms\\
    &$v_{\rm *}$ \tnote{d}& 203 & \kms\\
    &$\left<v_{\rm e}\right>\pm \sigma_{\rm e}$ &  $786_{-304}^{+450}$ & \kms\\
    \cline{2-4}
    &&&\\
    &\multicolumn{3}{c}{\textit{Accretion disk light profile}}\\
    &\rzero & [0.1 - 6.1]&  \rref\\
    &\rref \tnote{e}  \xspace \xspace & 1.62$\times 10^{15}$   & cm \\
    \hline
    \hline
    \multirow{7}{*}{\begin{sideways}Reverberation\end{sideways}}&&&\\
    &&&\\
    &\fblr \tnote{a} & 0.43$\pm$0.034 & - \\
    &\rblr & [0.1 - 2.5] & \rblrref\\
    &\rblrref \tnote{d}&1.71$\times 10^{17}$ & cm\\
    &&&\\
    &&&\\
\end{tabular}
\begin{tablenotes}\footnotesize
\item [a] see Sect.~\ref{results}
\item [b] taken from \citet{morgan08}
\item [c] \citet{morgan12}
\item [d] \citet{kogut93}
\item [e] \citet{mosquera11}
\end{tablenotes}
\end{threeparttable}

Eventually, the posterior probability is obtained using Bayes theorem:
\begin{equation}
   P(\zeta | d) = \mathcal{L}(d|\zeta) \cdot \mathcal{P}(\zeta)/E(d),
    \label{posterior}
\end{equation}
where $\mathcal{P}(\boldsymbol{\zeta}$) is the prior probability of the parameters $\boldsymbol{\zeta}$ and $E(d)$ the probability of the data, i.e. the Bayesian evidence. Calculating $E(d)$ requires the integration of the posterior across the whole parameter space, which is too demanding in computational time to perform here. This means that we cannot compare different models, but we can still use relative probabilities within any given model.
All parameters are assumed to have a uniform prior except \ve, whose prior is given in Eq.~(\ref{ve}).
Finally, we obtain the posterior probability marginalised over a given parameter or subset of parameter $\zeta_i$ through:
\begin{equation}
    P(\zeta_{\rm i} | d) = \int_{\rm j\neq i} P(\zeta_{\rm j}| d) \cdot \mathcal{P}(\zeta_{\rm j})d\zeta_{\rm j}.
    \label{eq:marginalisation}
\end{equation}

\begin{figure}[]
\centering
\includegraphics[width=0.5\textwidth]{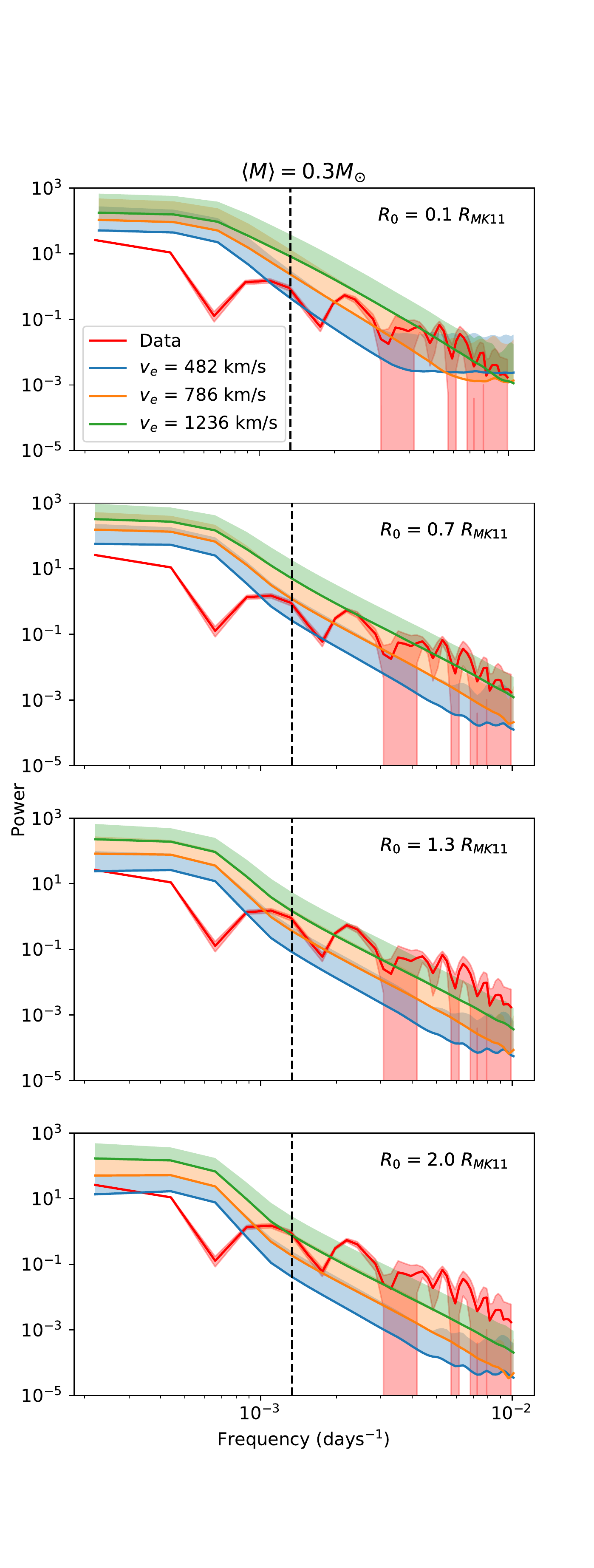}
\vspace{-2.2cm}
\caption{Mean (lines) and upper 1\tiretsigma envelope (shaded area) of power spectra from 100'000 simulated curves for different (\ve, \rzero\xspace) configurations, in absence of reverberation ($F_{\rm BLR}=0$ in Eq.~\ref{fsluse}), compared to the data (same as Fig. \ref{powerspectrum_data}). Due to the logarithmic scale, the lower envelopes extend almost to the x-axis and are not displayed for clarity. The vertical dashed line marks the low/high frequency boundary. \label{ex_ps}}
\end{figure}

\begin{figure*}[h]
\centering
\includegraphics[width=1.\textwidth]{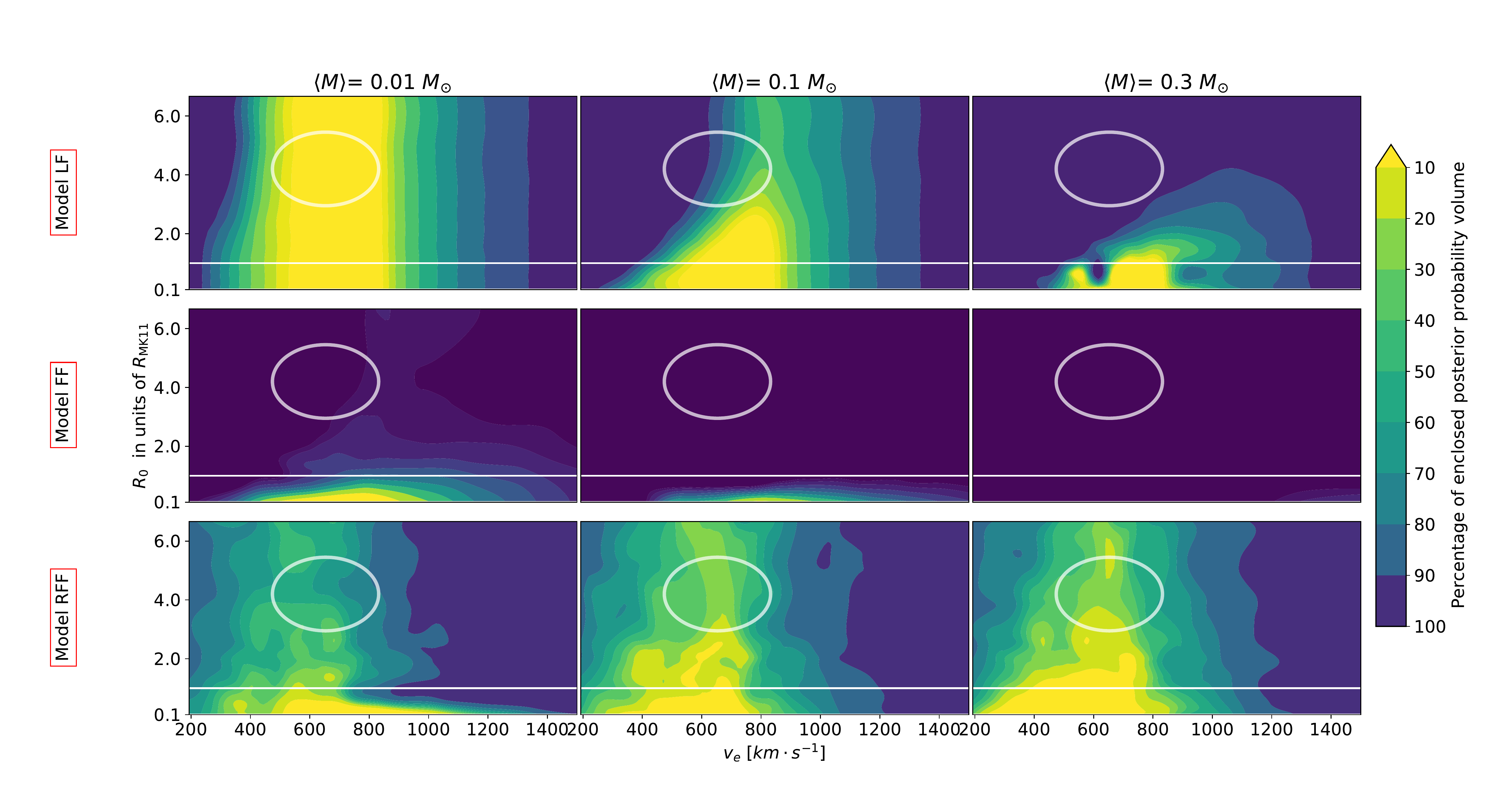}
\vspace{-1.0cm}
\caption{ Slices of the marginalised posterior probability of the (\ve, \rzero\xspace, \meanm) parameter space (three-dimensional after the marginalization over the angle of \ve using Eq.~\ref{eq:marginalisation}) for each of the three models described in Sect. \ref{results}.  \textit{Model LF}: the light curves are simulated without reverberation and fitted only to the low frequency data, i.e. up to 1/750 \invday, that corresponds to the same frequency cut as in \citet{morgan12}. \textit{Model FF}: the same simulations are fitted to the full frequency range, i.e. up to 1/100 \invday. \textit{Model RFF}: the light curves are now simulated with reverberation (\fblr = 0.432$\pm$ 0.036, \sigmadrw = 55
and \rblr = \rblrref) and the full observed frequency range up to 1/100 \invday is considered. The solid line corresponds to \rref, the estimate of \citet{mosquera11}, the ellipse represents the (\ve, \rzero\xspace) measurement interval from \citet{morgan12}. The coloured contours encapsulate [10-100]\% of the probability's volume and are projected in each of the displayed slices. We note that our probability densities are not scaled by the evidence and therefore cannot be compared across different models.
\label{res_mpg}}
\end{figure*}

\section{Results \label{results}}



We study the effect of high-frequency variability, like the one introduced by a reverberated BLR component, in measuring the size of the accretion disk.
In doing so, we come up with a new way to measure the size of the BLR by using the microlensing light curves from Eq.~(\ref{diff_ml}) in the full frequency range.
Before describing our results, we present our prior assumptions for the various model parameters, listed in Table \ref{param_space}.

\textit{Intrinsic variability:} The long term brightness decrease of image B (see Fig.~\ref{lc_0158}) compared to the behaviour of the A light curve, which consists of oscillations around a mean, suggests two possible scenarios: the intrinsic luminosity of the quasar is decreasing and image A is microlensed or the intrinsic luminosity of the quasar is rather constant and a microlensing event which started in image B before the beginning of the observations is now ending leading to a decrease of the micro-magnification. According to \citet{macleod16}, photometric changes of $|\Delta m|\geq 1\ $mag over $\approx$ 10 years as is observed in the B light curve, are very rare (around 1\% of quasars display this kind of variability). Furthermore, spectra of \Jzeroun shown in \citet{faure09} (i.e. taken during the first quarter of the light curve of Fig.~\ref{lc_0158}), show a typical Type 1 QSO spectrum for image A whereas the B spectrum shows faint and deformed emission lines. Altogether, these observation lead us to favour the second scenario. We will therefore consider image A to be microlensing-free, and therefore we use its light curve as a proxy for the quasar intrinsic variability. 
Using the \textsc{JAVELIN} software \citep{zu13} that employs a maximum likelihood approach in a Markov Chain Monte Carlo (MCMC) framework we find $\tau_{\rm DRW} =810$ days for the assumed damped random walk intrinsic variability model.
However, the observed light curve is the result of a convolution between the driving source and the light profile of the quasar. Therefore, the amplitude of the variations, \sigmadrw, cannot be constrained because it is degenerate with the radius of the source \rzero\xspace which is also unknown in this study. A large interval is hence considered for exploring this parameter that includes the values of \sigmadrw for all the known quasars \citep{ivezic20}.

\textit{Lens mass model and magnification maps}: \citet{morgan12} explore a list of lens mass models having a stellar mass fraction, \fml, between 0.1 and 1, and give a relation between \fml and the time delay between the two images $\Delta t$.
Using this relation and the time delay measured by \citet{millon20} we obtain \fml = 0.9, which we use throughout the following.
We adopt the $\kappa$, $\gamma$ and $\kappa_{\rm *}$ values at image locations from \citet{morgan08}, also listed in Table~\ref{param_space}, to compute the magnification maps that we will use. As model uncertainties are not given, we assume $\delta \kappa, \delta \gamma \leq 0.01$, quoted in most modelling works \citep[e.g. see Table B1 of][]{wong17}. According to \citet{vernardos14}, magnification maps within these uncertainties have statistically equivalent magnification probability distribution. As a sanity check, the same experiment was performed with a different mass model with $\Delta \kappa, \Delta \gamma \geq 0.03$ leading to the same broad conclusions. We therefore do not expect the uncertainty on the macro-model to influence on our study.

In most microlensing light curve fitting studies \citep{kochanek04, morgan08, cornachione19}, \meanm = 0.3 \solarm is taken as a reference mass around which a range of mean mass is explored.
Because we are interested in high-frequency variability, we also explore the effect of smaller values of the mean mass, viz. \meanm = 0.1 \solarm and \meanm = 0.01 \solarm, that can introduce shorter microlensing events for any given effective velocity due to the corresponding smaller physical size of the caustics.
Choosing a shallower Chabrier IMF, instead of the steeper Salpeter one used here, leads to fewer low mass microlenses and therefore reduces any effect of high-frequency variability.
Although this has been shown to affect magnification map properties \citep[see][]{chan20}, our goal here is to understand such short-scale variability and therefore we stick to the Salpeter IMF for all values of \meanm.

\textit{Accretion disk size:} In order to limit the number of free parameters of our study, we assume a face-on thin-disk model. The expected maximal inclination angle of a type 1 AGN is $\sim$60 degrees with respect to the line of sight \citep[e.g.][]{borguet08,poindexter10} and can induce, at most, a factor two systematic effect on the determination of \rzero. 
Nevertheless, \rzero, \ve, and the inclination angle are degenerate. We have therefore repeated our measurement of \rblr\, each time varying \rzero\, and \ve by factors of several and found no significant differences with our results in the face-on disk assumption. We use $\log(R_0/cm) = 15.07 \equiv \rref$ from \citet{mosquera11} as value of reference for the scale radius.
We explore \rzero\xspace in the range $0.1 - 6 \times$ \rref, which is bound at the low end by the magnification map resolution and extends high enough to include the measurement of \citet{morgan12}.

\textit{Reverberated variability:}  As mentioned previously, microlensing is dominant in image compared to image A.
Therefore we use image A's spectrum to derive \fblr in order to avoid any contamination from a possibly microlensed continuum.
To do so, the spectrum presented in \citet{faure09} was analysed using a multi-component decomposition as in \citet{sluse12}.
However, contrary to \citet{sluse12}, a MCMC approach was used to estimate the median \fblr and a 68\% credible interval (see Table \ref{param_space}). We emphasize the fact that \fblr is computed as the fraction of flux coming from the BLR compared to the continuum in the R-band, irrespective of the atomic species, therefore both the \ion{Mg}{ii} and \ion{Fe}{ii} emissions are included. 
As for the radius of the BLR, \citet{mosquera11} used the H$\beta$-BLR size-luminosity relationship \citep{bentz09} to estimate \rblr $= 1.71\times 10^{17}$cm $\approx 39$ light days, which we adopt here as our reference value, \rblrref.
The lower bound of the \rblr range that we explore is 0.1 $\times$ \rblrref $\approx$ 5 light days, i.e the smallest reverberation delay observable with the sampling of the light curve set to 1 point every 2-3 days.
The upper bound is set to 2.5 $\times$ \rblrref to include the confidence interval of the \citet{mosquera11} estimate.

To understand the effect of key parameters in the frequency of the signal in the simulated light curves we provide an illustrative example in Fig.~\ref{ex_ps}, where we show power spectra calculated for different values of the transverse velocity, \ve with corresponding directions drawn from the probability density function shown inf Fig.~\ref{vdistrib}, and the scale radius, \rzero\xspace, two of the main free parameters in subsequent models.
Firstly, we note that for a given value of \rzero\xspace the power has a tendency to increase with \ve, which is justified because the higher the velocity the faster the source is crossing caustics, inducing more high magnification events in both the high and low frequencies. 
Secondly, for a given value of \ve, the power in the high frequencies is inversely proportional to \rzero\xspace.
This is explained by the microlenses magnifying an ever decreasing portion of larger accretion disks, with the resulting magnification effect being diluted within the overall flux, leading to smoother and weaker high-frequency variations.

Furthermore, Fig.~\ref{ex_ps} shows that in the low frequency regime the models match reasonably well the data, while most of the model differences occur in the high frequencies. 
It is almost impossible to simultaneously match both the low and high frequencies, whatever the model parameters may be. This suggests that other physical mechanisms might be at play, in addition to microlensing, which we explore with the following three experiments:



\begin{itemize}
\item Low Frequency (LF): we apply the power spectrum method in the same setup as \citet{morgan12}, viz. we use only the low frequency part of the power spectrum, imposing a cutoff at 1/750 \invday that corresponds to the typical time-scale considered in Figure~2 of \citet[][]{morgan12}. At this stage, we do not include any reverberation signal and set $F_{\rm BLR}=0$ in Eq.~(\ref{fsluse}). As a result, even though intrinsic variability is included in this model, it is cancelled out since we are studying the differential microlensing light curves. This model is therefore not sensitive to the intrinsic variability parameters \sigmadrw and $\tau_{DRW}$.
\item Full Frequency (FF): we perform the same analysis as above, this time including the high frequencies up to 1/100 \invday.
\item Reverberation Full Frequency (RFF): we use the full frequency range of the data, as in the FF model, but this time we include the reverberation of the continuum.
\end{itemize}
The details and outcomes of each experimental set-up are detailed below.

\subsection{LF: low frequencies without reverberation \label{resmodel1}}

In \citet{morgan12}, the shortest features in the light curves last approximately two consecutive seasons, i.e. $\approx$750 days (see their Figure 2). This translates in frequencies of up to 1/750 \invday, which is lower than the 1/100 \invday limit that we have set in Sect. \ref{data}. We therefore adopt 1/750 \invday as being our boundary between what we define as low and high frequencies. Here, we model only the low frequency power spectrum of the data that contains by design the same signal frequencies as the data used in \citet{morgan12}.

In the top row of Fig.~\ref{res_mpg} we show the posterior probability from Eq.~(\ref{posterior}) as a function of \ve, \rzero, and \meanm.
The distribution of \rzero\xspace broadens with decreasing \meanm, becoming almost uniform for the smallest mean mass of \meanm = 0.01 \solarm, and therefore not providing any useful constraint.
This can be understood in terms of the physical size of the magnification maps, which depends on \meanm through the Einstein radius of the microlenses, \thetae (see Eq.~\ref{thetaE} and Table~\ref{param_space}), with respect to the velocity: decreasing \meanm is equivalent to rescaling the magnification map to a smaller physical size that allows the source to cross the map more rapidly for the same effective velocity (which is equivalent to increasing the effective velocity while keeping the mass fixed).
Thus, small masses and largest radii can induce enough high-frequency power to fit the data equally well as the smaller radii and larger masses.

Overall, our power spectrum measurement is in good agreement with the estimate of \citet{mosquera11}, while it is consistent within 1 to 2 $\sigma$ with the result of \citet{morgan12} in the \meanm = 0.01- 0.1\solarm cases. In the \meanm = 0.3 \solarm case we note a slight discrepancy with the results of \citet{morgan12}. This can be explained by our use of longer light curves (6 more seasons) and the use of a model driven prior on the angle of \ve as illustrated in Fig.~\ref{vdistrib}, instead of the uniform prior used in previous studies\footnote{The experiment ran with a uniform prior on the angle actually yields a 1 to 2\tiretsigma compatible measurement.}.  
Model LF is a sanity check demonstrating that, when restricted to low frequencies, the power spectrum and light curve fitting methods give compatible results and the data can be explained by microlensing alone.


\subsection{FF: full frequency range without reverberation \label{resmodel2}}
We now include the high-frequency signal in the data and attempt to explain it assuming that the observed variations come solely from microlensing of the accretion disk, i.e. exactly the same model as in the LF setup.
Our results shown in the second row of Fig.~\ref{res_mpg} favour much smaller accretion disks with \rzero$<0.5$ \rref, excluding both the result of \citet{morgan12} and the estimation of \citet{mosquera11}.
This is less prominent for the case with \meanm$=0.01$ \solarm, where we observe the same behaviour for larger \rzero\xspace as in the LF case, extending the compatible sizes to somewhat larger values.
Clearly, microlensing alone has problems in explaining the high-frequency signal and we need to invoke additional sources of variability to explain the data, as we do in the next case.

\begin{figure}[h]
\centering
\includegraphics[width=1.1\linewidth]{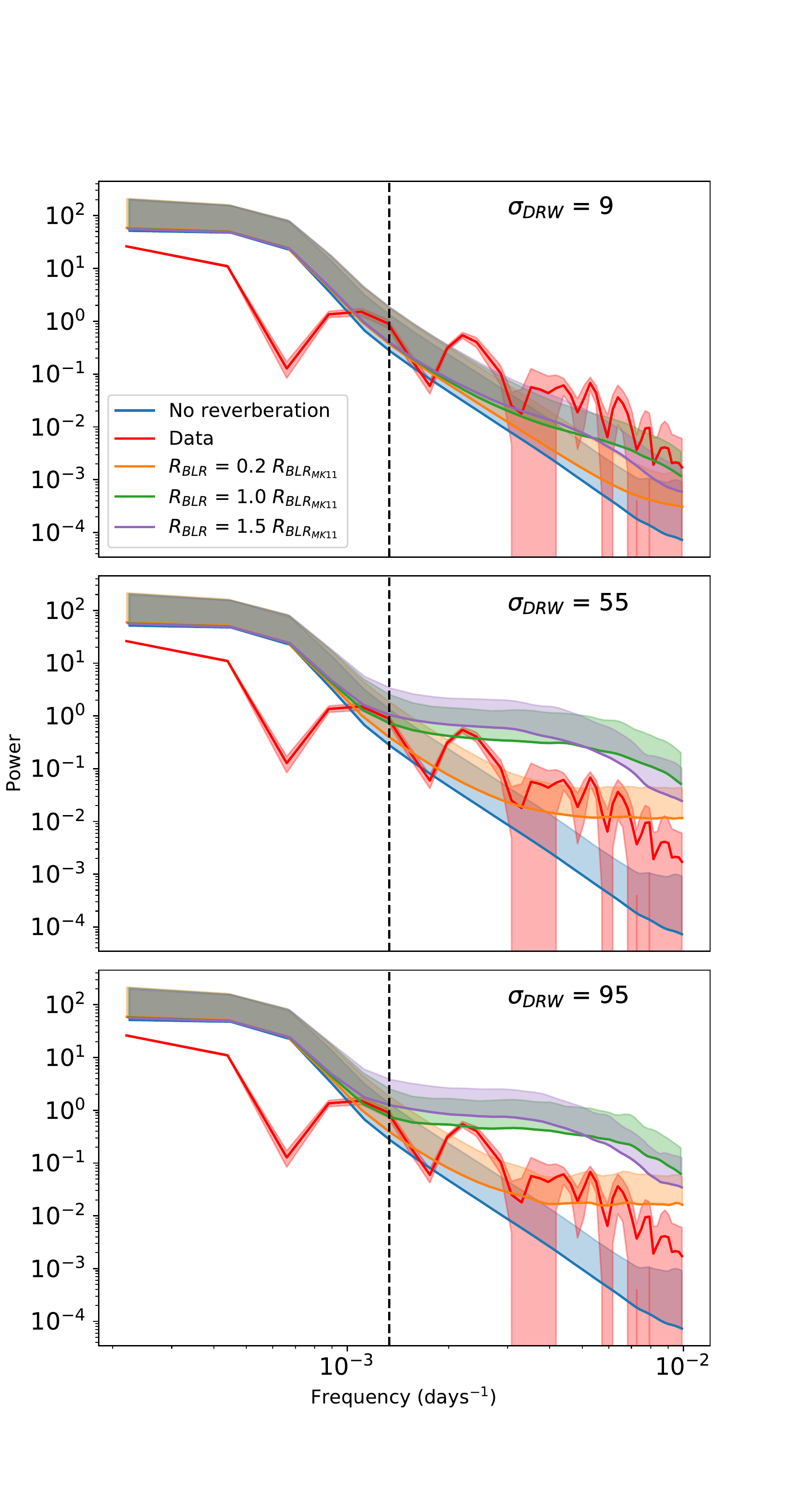}
\vspace{-1.5cm}
\caption{Effect of the reverberation process on the power spectrum. The data are shown as a solid red line. The coloured envelopes display the power spectra from 100'000 simulated curves for different values of \rblr. Each panel considers different values for \sigmadrw. In this plot we use \rzero = \rref, \meanm = 0.3 \solarm, \ve = 700 \kms $~and~$ \fblr=0.432 $\pm$ 0.036. The black dashed line marks the low/high-frequency limit. While the high-frequency range is never well represented with pure microlensing (blue), it is very sensitive to a change in the reverberation parameters. \label{mean_ps_reverberation}}
\end{figure}

\subsection{RFF: full frequency range with reverberation \label{resmodel3}}
\begin{figure*}[h]
\includegraphics[width=1.0\linewidth]{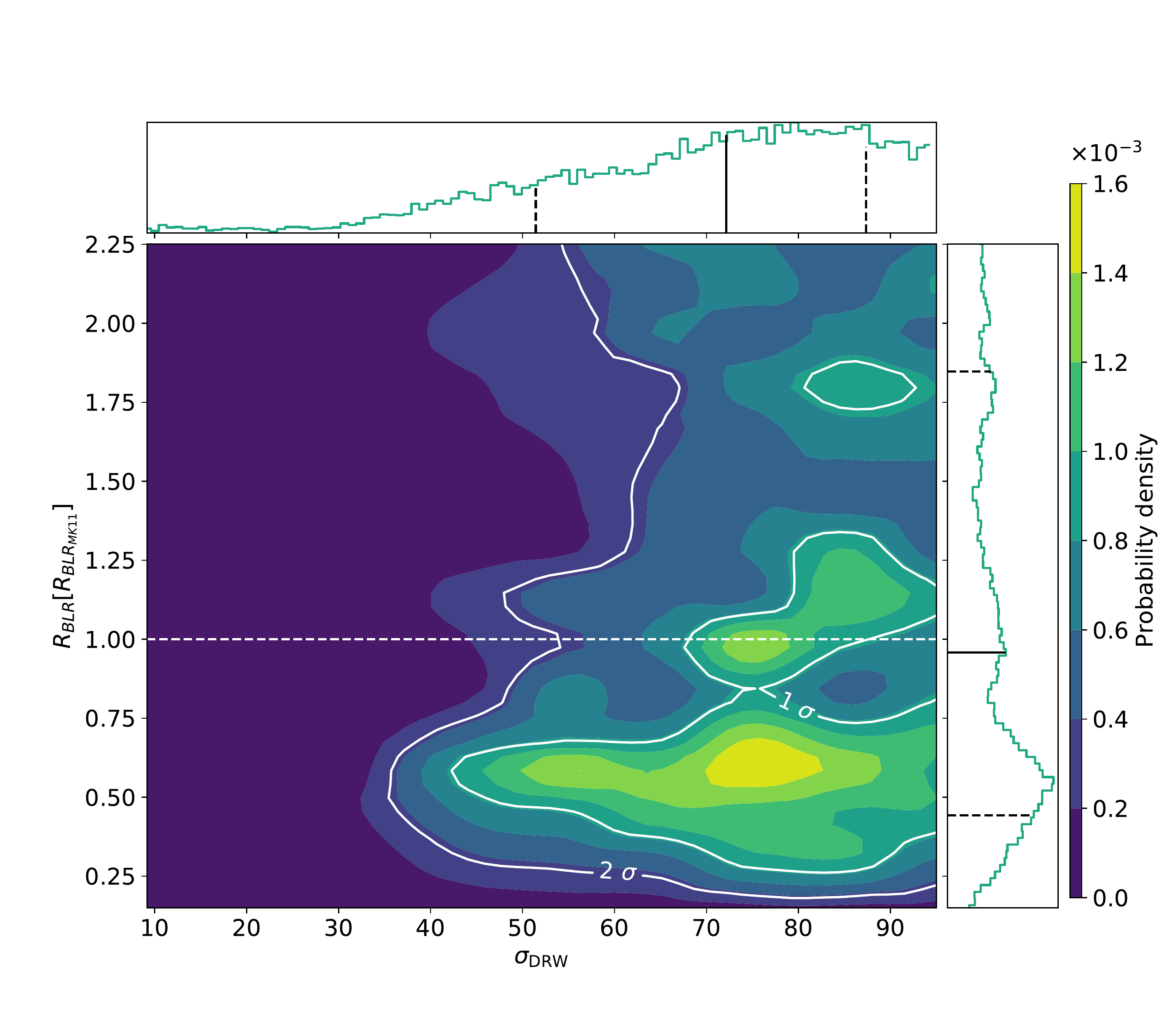}
\vspace{-1.2cm}
\caption{Posterior probability density in the (\sigmadrw, \rblr) parameter space marginalised over the microlensing parameters given \meanm = 0.3\solarm. \rblr is given in units of \rblrref=$1.71\times 10^{17}$cm, indicated by the white dashed line. The contours correspond to the 1 and 2\tiretsigma confidence intervals. Marginalized probability distributions of \sigmadrw and \rblr are given in the top and right histograms respectively. In each histogram, the black line shows the 50th percentile (median value) of the distribution and the dashed lines highlight the 16th and 84th percentile. Hence, we obtain \sigmadrw = $72^{+16}_{-21}$ and \rblr = $1.6^{+1.5}_{-0.8}\times 10^{17}$cm. \label{posterior_reverberation}}
\end{figure*}

Adding the reverberation process to the simulated light curves is expected to increase the power of the high frequencies in the signal.
In the bottom row of Fig.~\ref{res_mpg} we show the posterior probability as a function of  \ve, \rzero\xspace, and \meanm for a fiducial reverberation model with \rblr = \rblrref and \sigmadrw = 55 (see also Fig.~\ref{posterior_reverberation} and section \ref{sec:investigate_reverb} ).

As we can see on the third row of Fig.~\ref{res_mpg}, when including the reverberation effect, the constraint obtained for the radius of the accretion disk \rzero\xspace is now dominated by the prior on \ve and \rzero\xspace for every \meanm explored.  
Still, we can notice that the areas corresponding to any given percentage of enclosed posterior probability shrinks with the value of \meanm. Indeed, if a simulated power spectrum is compatible with the data for \meanm = 0.3 \solarm, the addition of power induced by the decrease of \meanm (as discussed in section \ref{resmodel1}) pulls the simulated power spectrum away from the data hence reducing their compatibility. Therefore, this model tends to favour the standard value of \meanm =0.3 \solarm.

\subsection{\rblr measurement \label{sec:investigate_reverb}}
In order to measure the size of the reverberating region, \rblr, we first explore the effect of the amplitude of the intrinsic variability, \sigmadrw, and of \rblr, on the simulated power spectra, while keeping the microlensing parameters fixed, i.e. \meanm = 0.3 \solarm, \rzero\xspace=\rref, and \ve = 700 \kms. The use of \meanm = 0.3 \solarm is motivated by the fact that galaxies are  unlikely to host a population of objects with \meanm = 0.1 or 0.01 \solarm (see Section~\ref{resmodel3}). Including reverberation in the analysis allows us to explain the high frequency using a realistic value of \meanm, or at least a consensus one. 
As for \ve, we use a value close to the mean value from Eq.~(\ref{ve}, see also Table \ref{param_space} and Fig. \ref{vdistrib}).  We stress the fact that, if the microlensing is identical in both images (i.e. $\mu_{A}(t)$=$\mu_{B}(t)$ in Eq.~\ref{fsluse}, which is more likely to happen if both images are not microlensed), the effect of reverberation is absent from the differential light curve we analyse. Therefore reverberation is not a stand-alone part of this study and the reverberation-induced variability ends up being weighted as a function of time because of microlensing in the differential light curve. 
Fig.~\ref{mean_ps_reverberation} shows that increasing \sigmadrw leads to more power at the high frequencies, mostly because the now stronger intrinsic variations are reverberated after a time lag ($\tau=$\rblr$/c$, see Eq.~\ref{transferfunction}) of the order of tens to hundreds of days, i.e. with a frequency $>1/750$ \invday.
The effect of the BLR size, \rblr, is more complex to analyse.
For a given transfer function, $\Psi(t, \rblr/c)$, a short variation of typically $\approx 100$ days in the intrinsic signal $I(t)$, appears twice in light curves simulated using Eq.~(\ref{fsluse}): once at time $t$ and a second time at $t+\rblr/c$.
As \rblr is increased, the echoed signal moves further eventually becoming fully separated spatially from the one originating at the disk, and is seen as a whole new feature of the light curve. 
This adds power to the high-frequency domain, justifying the difference between the \rblr = 0.2 \rblrref and \rblr = \rblrref cases in Fig. \ref{mean_ps_reverberation}.
As we keep increasing \rblr, $\Psi(t, \rblr/c)$ gets wider and starts to smooth out the short intrinsic variations, reducing their power.
This explains the drop in power in the highest of the frequencies when considering the highest values of \rblr in Fig. \ref{mean_ps_reverberation}. 

Comparing Figs. \ref{ex_ps} and \ref{mean_ps_reverberation}, we see that a reverberated variability component has a stronger effect on the high-frequency power than increasing \ve or decreasing \rzero\xspace.
This is to be expected because intrinsic quasar variations generally have shorter time scales than microlensing.
As a consequence the choice of microlensing model (i.e the \ve and \rzero\xspace values) has little impact on the measurement of \rblr.

Using Eq.~(\ref{posterior}), we derive the probability density in the parameter space (\sigmadrw, \rblr) for a given set of parameters \meanm, \fblr, \ve, and \rzero\xspace. By marginalizing on \ve and \rzero\xspace we obtain the measurement of \rblr shown in Fig.~\ref{posterior_reverberation}. One could argue that we obtain a bi-modal distribution in the posterior probability for \rblr. This observation can be explained by the fact that, as shown in Sect.~\ref{results}, the R-band encapsulates the \ion{Mg}{ii} and \ion{Fe}{ii} emission lines which can arise from two distinct regions of the BLR. Indeed, the H$\beta$ (used in \citet{mosquera11}) and \ion{Mg}{ii} lines seem to arise from the same part of the BLR in various quasars \citep[e.g.][]{kaouzos15, khadka21} and should both yield similar sizes; whereas the \ion{Fe}{ii} line is thought to arise from a larger part of the BLR \citep[e.g.][]{sluse07,hu15,zhang19,li21}. Therefore, the combination of the two signals modelled as a single BLR emission could broaden our measurement and induce its slight bi-modality. Still, the core of the probability lies in the [0.1-1.5]\rblrref range and the second mode observed for higher values of \rblr rises only for the highest values of \sigmadrw. The marginalization of this posterior over \sigmadrw yields a probability distribution for \rblr and by taking its 16th, 50th and 84th percentiles we measure \rblr = $1.6^{+1.5}_{-0.8}\times 10^{17}$cm. With a relative precision of $\approx 80\%$ our method is less precise than recent spectroscopical reverberation mapping measurements \citep[e.g.][ have around 30\% relative precision for quasars with z>1.3]{grier19,penton21} but more precise than photometric reverberation mapping \citep[e.g.][have above 100\% relative precision when using cross correlation function with R and B filter light curves]{Kaspi21}. The value of \rblrref predicted by the luminosity-size relation is in agreement with our measurement at the 1\tiretsigma level. 



\section{Discussion \label{discussion}}

We now review the implications of the constraints on the accretion disk scale radius, \rzero\xspace, found using the three different models. 
The first model shows that the low frequency variations of the microlensing light curve do not have a strong constraining power on \rzero\xspace when using the power spectrum method. 
The second model indicates that the high-frequency part of the power spectrum adds significant constraints to the accretion disk measurement since the range of \rzero\xspace compatible with the data is shrunk and leans towards the smallest values. Therefore, ignoring the high-frequency variations may lead to an overestimation of \rzero\xspace. 

The first two models which rely only on microlensing variability both require lower values of the mean stellar mass \meanm. However, a galaxy populated by stars with \meanm = 0.01\solarm is hardly conceivable since the least massive star known to this day has a mass of 0.07\solarm \citep{kasper07}. This means that, according to these two models, the population of compact objects that is most likely to produce the observed variability, is not made of stars. Hypothetical populations of primordial black holes \citep{hawkins20a, hawkins20b} and galaxies with significant part of brown dwarfs and/or free-floating Jupiter-like planets \citep{dai18, cornachione20} have been invoked to explain unexpected microlensing features. Unfortunately, primordial black holes have, to date, never been observed in nearby galaxies despite huge efforts of multiple collaborations \citep{alcock01, niikura19a, niikura19b}. In addition, their theoretical mass is poorly constrained and spans the very broad range $\left[10^{-16}, 10^2\right]$\solarm \citep{green20}. Similarly, a low-mass stellar population is not observed in the Milky Way \citep{mroz17}. In both cases, the explanation behind the observed variability relies on an exotic population of microlenses in the lens galaxy for which we do not have any observational proof so far.

Continuum reverberation in the BLR is an acknowledged and observed effect \citep{blandford82,bentz09,du16,williams20} and supports the validity of our third model. The latter encapsulates our best understanding of microlensing light curve variability. It is also the only model that favours a more standard value of the mean stellar mass, \meanm = 0.3\solarm, not requiring any exotic population of microlenses. This suggests that reverberation of the continuum by the BLR, which was observed multiple times with spectroscopic monitoring, is also observable in single-band photometric light curves through their high-frequency variations on the differential light curve.

Last but not least, this offers a new way of measuring the size of the BLR illustrated by Fig.~\ref{posterior_reverberation}. The relative insensitivity of this measurement to the microlensing parameters (\ve, \rzero\xspace) is due to the fact that, as stated in Sect. \ref{resmodel1}, the main challenge of a given set of parameters $\zeta$ is to fit the high-frequency power and these are mainly set by the reverberated variability. The downside of this is that we are not able to discriminate between the values of \rzero\xspace. Nevertheless, in the case of \Jzeroun the measurement of \rblr is in agreement with the H$\beta$-BLR size-luminosity relation \citep{mosquera11}.

\section{Conclusions and Perspectives}
In this work, we present a new method employing the power spectrum of microlensing light curves to study the strongly lensed quasar \Jzeroun system. This method allows us to take into account both the high-frequency variations of the data and include the reverberated variability in the microlensing paradigm. Our main results are summed up in the following points:
\begin{enumerate}
    \item Ignoring the high-frequency, as it is the case with the light curve fitting method, may lead to an overestimation of the scale radius of the accretion disk \rzero\xspace. In fact, we show that the use of short variations excludes the values of \rzero\xspace found with the light curve fitting method in \citet{morgan12}.
    \item In the context of standard paradigm of microlensing light curve simulation, the data favour an exotic microlens population drawn from an IMF with unprecedented low mean mass \meanm whereas with the model we propose, including the BLR reverberation, the data favour stellar populations drawn from a standard Salpeter IMF with \meanm = 0.3 \solarm.  
    \item For the first time, continuum reverberation by the BLR is observed in a single waveband photometric light curve. We use this opportunity to measure the size of the BLR in \Jzeroun to be \rblr = $1.6^{+1.5}_{-0.8}\times 10^{17}$cm, compatible with the expectation of the luminosity-size relation with a better precision than standard photometric reverberation mapping techniques.   
    \item The power spectrum fitting method is insensitive to the scale radius of the accretion disk \rzero\xspace in presence of reverberated variability in the single waveband light curve. 
\end{enumerate}
With the encouraging results this method gave with \Jzeroun, we are looking forward to applying it on other systems for which a microlensing light curve is available. In the upcoming LSST era, this method offers a new way to probe the luminosity-BLR size relation of quasars for a large range of redshifts and luminosities.

\begin{acknowledgements}
This work is supported by the Swiss National Science Foundation (SNSF) and by the European Research Council (ERC) under the European Union's Horizon 2020 research and innovation program (COSMICLENS: grant agreement No 787886). GV has received funding from the European Union's Horizon 2020 research and innovation program under the Marie Sklodovska-Curie grant agreement No 897124. The authors wish to thank Aymeric Galan for useful comments and graphical contribution. Finally, we thank the anonymous referee for the useful comments that improved the clarity of the paper.
\end{acknowledgements}

\bibliographystyle{aa}
\bibliography{pdmbib}
\clearpage

\end{document}